\documentclass{sig-alternate-05-2015}

\usepackage{tikz}
\usepackage{pgfplots}
\pgfplotsset{compat=newest} 
\pgfplotsset{plot coordinates/math parser=false} 
\newlength\fheight
\newlength\fwidth
\definecolor{chaptergrey}{rgb}{0.61, 0, 0.09} % dialed back a little
\usetikzlibrary{calc}
\definecolor{chaptergreen}{rgb}{0.09, 0.612, 0}
\definecolor{chapterpurple}{rgb}{0.522, 0, 0.612}
\definecolor{chapterlightgreen}{rgb}{0, 0.612, 0.522}
\usepackage{subcaption}
\usepackage{booktabs}

\begin{document}

% tikz styles
\tikzstyle{startstop} = [rectangle, rounded corners, minimum width=2cm, minimum height=0.5cm,text centered, draw=black]
\tikzstyle{io} = [trapezium, trapezium left angle=70, trapezium right angle=110, minimum width=3cm, minimum height=1cm, text centered, draw=black]
\tikzstyle{process} = [rectangle, minimum width=2cm, minimum height=0.5cm, text centered, draw=black, align=center]
\tikzstyle{decision} = [ellipse, minimum width=2cm, minimum height=1cm, text centered, draw=black]
\tikzstyle{arrow} = [thick,<->,>=stealth]
\tikzstyle{darrow} = [thick,<->,>=stealth,dashed]
\tikzstyle{sarrow} = [thick,->,>=stealth,dashed]
\usetikzlibrary{backgrounds}

% avoid footnote split
\interfootnotelinepenalty=10000

% Copyright
%\setcopyright{acmcopyright}
%\setcopyright{acmlicensed}
\setcopyright{rightsretained}
%\setcopyright{usgov}
%\setcopyright{usgovmixed}
%\setcopyright{cagov}
%\setcopyright{cagovmixed}

\title{Performance Comparison of Dual Connectivity and \\ Hard Handover for LTE-5G Tight Integration}
%\subtitle{[Extended Abstract]
%\titlenote{A full version of this paper is available as
%\textit{Author's Guide to Preparing ACM SIG Proceedings Using
%\LaTeX$2_\epsilon$\ and BibTeX} at

\numberofauthors{3} 
\author{Michele Polese$^*$, Marco Mezzavilla$^\dagger$, Michele Zorzi$^*$\\ \\
$^*$Department of Information Engineering, University of Padova, Italy \\
 e-mail: \{polesemi, zorzi\}@dei.unipd.it\\
$^\dagger$Tandon School of Engineering, New York University, Brooklyn NY, USA 
\\ e-mail: mezzavilla@nyu.edu}

\maketitle
\begin{abstract}
Communications at frequencies above 10 GHz (the mmWave band) are expected to play a major role for the next generation of cellular networks (5G), because of the potential multi-gigabit, ultra-low latency performance of this technology. mmWave frequencies however suffer from very high isotropic pathloss, which may result in cells with a much smaller coverage area than current LTE macrocells. High directionality techniques will be used to improve signal quality and extend coverage area, along with a high density deployment of mmWave base stations (BS). However, when propagation conditions are hard and it is difficult to provide high quality coverage with mmWave BS, it is necessary to rely on previous generation LTE base stations, which make use of lower frequencies (900 MHz - 3.5 GHz), which are less sensitive to blockage and experience lower pathloss. In order to provide ultra-reliable services to mobile users there is a need for network architectures that tightly and seamlessly integrate the LTE and mmWave Radio Access Technologies. In this paper we will present two possible alternatives for this integration and show how simulation tools can be used to assess and compare their performance.
\end{abstract}

\begin{picture}(0,-110)(0,-430)
\put(0,0){
\put(0,0){\footnotesize This paper was accepted for presentation at the ninth EAI} 
\put(0,-10){\footnotesize SIMUtools 2016 conference, August 22 - 23, 2016, Prague,}
\put(0,-20){\footnotesize Czech Republic.}}
\end{picture}

\begin{CCSXML}
<ccs2012>
<concept>
<concept_id>10003033.10003079.10003081</concept_id>
<concept_desc>Networks~Network simulations</concept_desc>
<concept_significance>500</concept_significance>
</concept>
<concept>
<concept_id>10003033.10003106.10003113</concept_id>
<concept_desc>Networks~Mobile networks</concept_desc>
<concept_significance>300</concept_significance>
</concept>
</ccs2012>
\end{CCSXML}

\ccsdesc[500]{Networks~Network simulations}
\ccsdesc[300]{Networks~Mobile networks}


%
%  Use this command to print the description
%
\printccsdesc

% We no longer use \terms command

\keywords{5G LTE tight integration, dual connectivity, simulation}

\section{Introduction}
The latest Cisco Visual Networking Index forecasts an explosion of mobile network traffic to be expected before 2020, with an eightfold increase of global mobile traffic~\cite{cisco}. The next generation of cellular networks will be required to address traffic demands, new business cases, and growth of connected devices, and to provide improved performance for many different use cases. For example, 5G networks should support \cite{ngmn5g} (i) a rate of 50 Mbits or more even at cell edges, with peaks of 1 Gbps in particular cases; (ii) ultra-low end-to-end latency below 10 ms, with a stricter requirement of 1 ms for applications such as tactile internet; (iii) ultra-reliable communications, with ultra-high availability and reliability; (iv) massive Machine Type Communications (MTC) with very low power.

MmWave technology may play a major role in allowing 5G networks to meet these requirements. The spectrum above 10 GHz is not as fragmented as the microWave spectrum, therefore large contiguous chunks of bandwidth are available. However, there are several issues that must be studied in order to make mmWave technology well understood and market-ready. MmWave frequencies suffer from high isotropic pathloss, which can be compensated with massive MIMO and beamforming, thanks to the small size of the antennas, but also from blockages from solid materials (i.e., buildings of bricks and mortar) \cite{rappaport2, rappaport1}.

These extreme propagation conditions offer new design requirements for the PHY and MAC layers \cite{zorzimac}, however they also have consequences on the higher layers of the protocol stack. In particular, in order to provide an ultra-reliable service, user equipments (UEs) at mmWave frequencies will have to perform handover between neighboring mmWave base stations or to fall back to the legacy LTE network. The LTE base stations may offer a reliable backup in case the link with mmWave BS is disrupted and outage is experienced. However, the current LTE handover procedure may be too slow, and may result in service interruptions and excessive signaling. Therefore, we propose a dual connectivity (DC) solution, in which the UE is connected to both Radio Access Technologies (RATs), and performs fast switching between the two with a single RRC message. The dual connectivity is part of a set of proposals on tight integration between LTE and mmWave RATs, which can be found in \cite{dasilva}. 

A thorough simulation campaign on dual connectivity or handover between different RATs has however not yet been performed, since current simulation tools for LTE and 5G are not integrated. In this paper we will describe the work done to implement a simulator that offers the capabilities of using PHY and MAC layers of LTE and 5G in the same UE. It is based on the open source simulator \textit{network simulator 3} (ns--3, \cite{ns3}), and uses the 5G mmWave protocol stack developed by NYU \cite{mmWaveSim} together with the LTE module of ns--3 \cite{lena}.

The paper is organized as follows. In Sec.~\ref{sec:integration} we will describe two different ways to integrate LTE and mmWave networks. Then in Sec.~\ref{sec:implementation} we will present the simulation framework and briefly recall some implementation issues. Finally we will show which are the metrics we expect to measure from still ongoing simulation campaigns and present some early results in Sec.~\ref{sec:metrics}.

\section{LTE-5G tight integration}\label{sec:integration}
LTE and 5G are expected to integrate at a certain layer and to rely on a common core network infrastructure, following an idea of \textit{harmonization} strongly advocated by several stakeholders \cite{nokia, metis, dasilva}. By exploiting existing infrastructure and architectures, the deployment of 5G networks can be faster and requires lower capital expenditure (CAPEX). Mature elements of legacy networks are expected to be used also in 5G networks, and one of the areas that can benefit from this integration is reliability, due to the wide-area coverage of LTE macrocells. The discussion on which is the appropriate layer at which LTE and 5G should integrate is still ongoing. The idea is that from this \textit{integration} layer upwards the two stacks will have common protocols. The integration layer has an interface to lower layer primitives and functions whose implementation differs in the two radio access technologies. A possible integration may be placed at the MAC layer, exploiting the fact that probably PHY layer will be OFDM based, but this would limit the optimization that can be carried out on the design of mmWave MAC layers. The same problem presents for integration at the \textit{Radio Link Control} (RLC) layer, which is tightly coupled with the MAC layer by the scheduling decisions.
A more realistic integration can be performed at the \textit{Packet Data Convergence Protocol} (PDCP) layer, which is not strictly dependent on lower layers' functionalities. In this work we assume that the integration is performed at this layer.

Currently, a typical UE supports more than one possible RATs, i.e., it can use GSM, UMTS or LTE networks. However, these three technologies are not integrated and are very different from one another, both in the radio access and in the core network. In order to switch from LTE to one of the two other legacy networks, an involved inter RAT procedure is required. However, for LTE and 5G networks a new inter RAT handover (HO) that requires less signaling may be designed. A good starting point to lower the burden and the complexity of the HO may be to consider mmWave cells as LTE cells, and use the faster intra RAT handover procedure. This may be made possible by the control layer integration between LTE and 5G, which is part of the proposals around tight integration \cite{dasilva}.

% description of intra RAT HO?

However, even the intra RAT handover may not be a good solution for a seamless and ultra-fast switch between the two air interfaces. Indeed, it involves signaling both between the two BSs, on the X2 interface, and at the RRC layer, to communicate to the UE the information on the target cell. It also needs a non contention based random access procedure, and a message exchange with the \textit{Mobile Management Entity} (MME) in the core network in order to switch the path from the \textit{Serving Gateway} (SGW) to the BS to which the UE is performing the handover. 
The processing in the MME may require a one-way latency in the order of 10 ms, and during this interval packets must be forwarded from the BS from which the handover originated to the target BS. Moreover, a handover from a mmWave BS to an LTE cell should be triggered only if the mmWave link is going to be in outage, in order to remain connected to the mmWave cell as long as possible and benefit from its high throughput and low latency. In this case, the control signaling may experience some latency due to packets received with errors and retransmissions, or worse, the UE may not manage to transmit/receive the required messages at all, thereby detecting a Radio Link Failure (RLF). During the random access procedure, moreover, the UE is not capable of transmitting and receiving data, which is buffered at the BS and experiences an increased latency.

A different and more robust solution can be obtained using a dual connectivity (DC) setup, introduced here as an extension of 3GPP's LTE DC proposal~\cite{36842}. In the proposed scheme, the UE is simultaneously connected to both LTE and mmWave BSs. The LTE cell acts as a backup: since the UE is already connected, when the signal quality of the mmWave link degrades, there is no need to perform a complete handover with a random access procedure, to send path switching messages to the core network: a single RRC control message from the LTE BS to the UE is enough. 
The LTE cell is also a coordinator of all the neighboring mmWave cells, as proposed in \cite{giordani}, and an access point to the core network for dual connected devices. 
In this setup the UE transmits pilots that mmWave base stations can use to estimate the link quality (for example by measuring the SNR). These measurements are periodically sent from each mmWave cell to the LTE cell, via the X2 interface. Therefore, the LTE BS knows which is the best mmWave cell for each UE: notice that this reporting is forwarded to the closest LTE eNB, i.e., the one that may offer fallback once all mmWave cells are in outage. Once the UE is within coverage of an LTE cell, it performs a random access to the LTE \textit{evolved Node Base} (eNB) using the standardized LTE procedure. As soon as the LTE BS knows that the UE has successfully completed the access procedure, it instructs the UE to connect also to the mmWave eNB with the best SNR.  

In a dual connected setup, in the LTE eNB there is a PDCP and related RLC instance for each bearer, while in the mmWave eNB there is only an RLC instance. Then, data is forwarded from the core network to the LTE eNB, where the PDCP layer decides whether to send it via LTE RAT, or to forward it to the remote RLC layer in the mmWave eNB. The mmWave link is used unless in outage, in this case a switch signal is sent to the UE which starts receiving data on the LTE radio interface. Notice that this scheme allows also to decouple uplink and downlink connections. If the LTE cell detects that another mmWave eNB has a better signal quality than the current one, it may trigger and orchestrate a handover between the two mmWave cells, while the data transmission is switched to LTE.

The PDCP forwards data via the X2 interface, thus, since the mmWave eNB is usually preferred to LTE, and according to the latency of the connection between BSs, the PDCP layer and anchor point to the core network may be moved to the mmWave eNB to decrease the end to end latency of data transmissions.

\section{Simulation of LTE-5G tight integration in ns--3}\label{sec:implementation}
In order to assess the performance of both schemes, we integrated two already developed tools for LTE and 5G simulations. The first is the LTE module of the ns--3 simulator, whose upper layers are shared by the ns--3 mmWave simulator developed by NYU~\cite{mmWaveSim}. This tool has a very detailed characterization of the physical layer, which exploits real measurements in order to model small scale fading, beamforming and channel conditions at 28 and 73 GHz carrier frequencies~\cite{phy5g}. This simulator assumes a TDD physical layer, and the physical frame structure can be configured to support different OFDM numerologies. 

Even though the mmWave simulator uses the LTE higher layers, prior to this work it was not possible to set up a simulation with devices capable of both LTE and mmWave communications in the same scenario. Therefore, the core of the DC implementation is a new extension of the ns--3 \texttt{NetDevice}, called \texttt{McUeNetDevice}, which has a double stack: mmWave physical, MAC and \textit{Radio Resource Control} (RRC) layers, LTE physical, MAC, RRC layers, and a single UE \textit{Non Access Stratum} (NAS) layer. The latter has an interface to both RRC entities and is in charge of the exchange of information between them. 

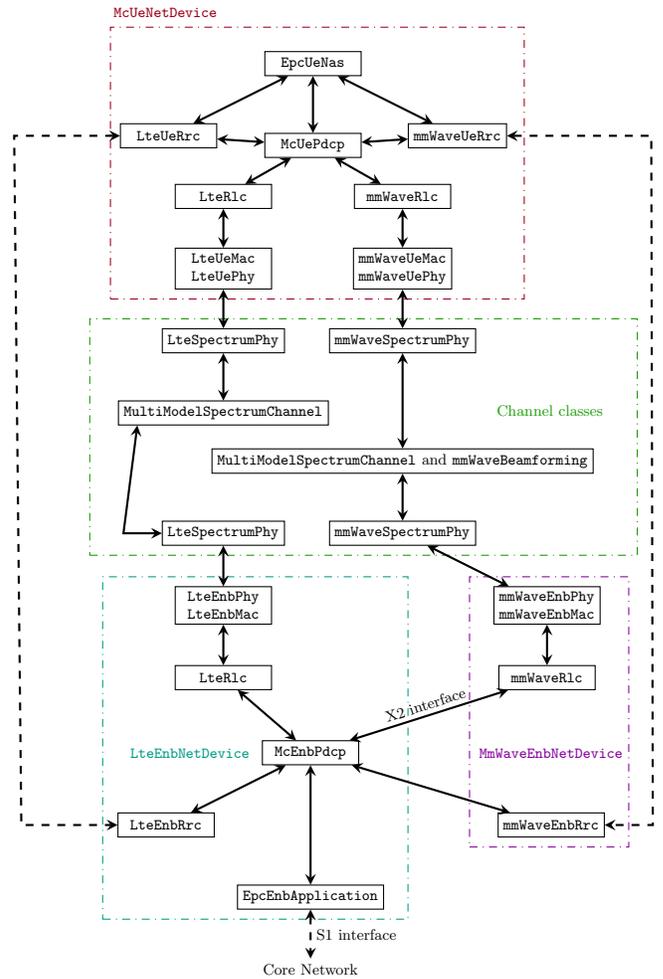
\begin{figure}[ht!]
\centering
\begin{tikzpicture}[node distance=1.5cm, scale=0.64, every node/.style={scale=0.64}]
  \node (epc) [process] {\texttt{EpcUeNas}};
  \node (rrc) [process, below of=epc, xshift=-3cm] {\texttt{LteUeRrc}};
  \node (mmrrc) [process, below of=epc, xshift=3cm] {\texttt{mmWaveUeRrc}};
  \node (pdcp) [process, below of=epc, yshift=-0.2cm] {\texttt{McUePdcp}};
  \node (rlc) [process, below left of=pdcp, xshift=-0.8cm] {\texttt{LteRlc}};
  \node (mmrlc) [process, below right of=pdcp, xshift=0.8cm] {\texttt{mmWaveRlc}};
  \node (ltephy) [process, below of=rlc] {\texttt{LteUeMac} \\ \texttt{LteUePhy}};
  \node (mmphy) [process, below of=mmrlc] {\texttt{mmWaveUeMac} \\ \texttt{mmWaveUePhy}};

  \draw[chaptergrey,dashdotted] ($(rrc.north west)+(-0.2,2)$) rectangle ($(mmphy.south east)+(1.5,-0.2)$);
  \node (legendMcUe) [above left of=epc, xshift=-2cm]{\textcolor{chaptergrey}{\texttt{McUeNetDevice}}};

  \node (ltephy2) [process, below of=ltephy] {\texttt{LteSpectrumPhy}};
  \node (mmphy2) [process, below of=mmphy] {\texttt{mmWaveSpectrumPhy}};
  \node (channel) [process, below of=mmphy2, yshift=-1cm] {\texttt{MultiModelSpectrumChannel} and \texttt{mmWaveBeamforming}};
  \node (ltechannel) [process, below of=ltephy2] {\texttt{MultiModelSpectrumChannel}};
  \node (enbltephy2) [process, below of=ltechannel, yshift = -1cm] {\texttt{LteSpectrumPhy}};
  \node (enbmmphy2) [process, below of=channel] {\texttt{mmWaveSpectrumPhy}};

  \draw[chaptergreen,dashdotted] ($(ltephy2.north west)+(-1.5,0.2)$) rectangle ($(enbmmphy2.south east)+(3.35,-0.2)$);
  \node (legendChannel) [above right of=channel, xshift=2cm] {\textcolor{chaptergreen}{Channel classes}};

  \node (lteEnbLl) [process, below of=enbltephy2] {\texttt{LteEnbPhy} \\ \texttt{LteEnbMac}};
  \node (mmEnbLl) [process, below of=enbmmphy2, xshift=3cm] {\texttt{mmWaveEnbPhy} \\ \texttt{mmWaveEnbMac}};
  \node (lteEnbRlc) [process, below of=lteEnbLl] {\texttt{LteRlc}};
  \node (mmEnbRlc) [process, below of=mmEnbLl] {\texttt{mmWaveRlc}};
  \node (enbPdcp) [process, below right of=lteEnbRlc, xshift=0.75cm, yshift=-0.5cm] {\texttt{McEnbPdcp}};
  \node (enbrrc) [process, below of=enbPdcp, xshift=-3cm] {\texttt{LteEnbRrc}};
  \node (enbmmrrc) [process, below of=enbPdcp, xshift=5cm] {\texttt{mmWaveEnbRrc}};
  \node (epcEnb) [process, below of=enbPdcp,yshift=-1.5cm] {\texttt{EpcEnbApplication}};
  \node (cn) [below of=epcEnb] {Core Network};

  \draw[chapterpurple,dashdotted] ($(mmEnbLl.north west)+(-0.5,0.2)$) rectangle ($(enbmmrrc.south east)+(0.5,-0.2)$);
  \node (legendChannel) [left of=enbPdcp, xshift=6.5cm] {\textcolor{chapterpurple}{\texttt{MmWaveEnbNetDevice}}};

  \draw[chapterlightgreen,dashdotted] ($(lteEnbLl.north west)+(-1.5,0.2)$) rectangle ($(epcEnb.south east)+(0.5,-0.2)$);
  \node (legendChannel) [left of=enbPdcp, xshift=-1cm] {\textcolor{chapterlightgreen}{\texttt{LteEnbNetDevice}}};

  % \node (ltephy) [process] {\texttt{LteUePhy}};
  % \node (mmphy) [process, right of=start, align=center] {Open socket, initialize\\useful variable, open input file};
  % \node (filesetup) [process, below of=setup, align=center, yshift=-1cm] {Compute file size\\ number of blocks $B$\\set number of tx block $b = 0$};
  % \node (while) [decision, below of=filesetup, yshift=-1cm] {$b < B$?};
  % \node (stop) [startstop, below left of=while, yshift=-1cm, xshift=-2cm] {Stop};
  % \node (read) [process, below right of=while, yshift=-1cm, xshift=1.2cm, align=center] {Read data, set\\packets needed $P = N$};
  % \node (send) [process, below of=read] {Encode and send $P$ packets};
  % \node (waitack) [process, left of=send, xshift=-4cm] {Wait for ACK};
  % \node (ack) [decision, below of=send, align=center, yshift=-1cm] {Packet needed\\$P$ = 0?};
  % \node (ackrx) [process, left of=ack, xshift=-4cm, align=center] {Update $P$};

  \draw[arrow] (epc) -- (rrc);
  \draw[arrow] (epc) -- (mmrrc);
  \draw[arrow] (epc) -- (pdcp);
  \draw[arrow] (mmrrc) --  (pdcp);
  \draw[arrow] (rrc) --  (pdcp);
  \draw[arrow] (pdcp) -- (rlc);
  \draw[arrow] (pdcp) -- (mmrlc);
  \draw[arrow] (rlc) -- (ltephy);
  \draw[arrow] (mmrlc) -- (mmphy);
  \draw[arrow] (mmphy) -- (mmphy2);
  \draw[arrow] (ltephy) -- (ltephy2);
  \draw[arrow] (ltephy2) -- (ltechannel);
  \draw[arrow] (mmphy2) -- (channel);
  \draw[arrow] (channel) -- (enbmmphy2);
  \draw[arrow] ([xshift=0.4cm]ltechannel.south west) -- node[anchor=east] {} ([xshift=-0.8cm]enbltephy2.west) -- (enbltephy2.west);
  \draw[arrow] (enbmmphy2) -- (mmEnbLl);
  \draw[arrow] (enbltephy2) -- (lteEnbLl);
  \draw[arrow] (mmEnbLl) -- (mmEnbRlc);
  \draw[arrow] (lteEnbLl) -- (lteEnbRlc);
  \draw[arrow] (lteEnbRlc) -- (enbPdcp);
  \draw[arrow] (mmEnbRlc) -- node[sloped, anchor=center, above] {X2 interface} (enbPdcp);
  \draw[arrow] (enbmmrrc) --  (enbPdcp);
  \draw[arrow] (enbrrc) --  (enbPdcp);
  \draw[arrow] (enbPdcp) -- (epcEnb);
  \draw[darrow] (rrc.west) -- node[anchor=east] {} ([xshift=-2.2cm]rrc.west) -- node[anchor=east] {} ([xshift=-2cm]enbrrc.west) -- (enbrrc.west);
  \draw[darrow] (mmrrc.east) -- node[anchor=east] {} ([xshift=+3cm]mmrrc.east) -- node[anchor=east] {} ([xshift=+1cm]enbmmrrc.east) -- (enbmmrrc.east);
  \draw[darrow] (epcEnb) -- node[anchor=west] {S1 interface} (cn);

  % \draw[arrow] (start) -- (setup);
  % \draw[arrow] (setup) -- (filesetup);
  % \draw[arrow] (filesetup) -- (while);
  % \draw[arrow] (while) -- node[anchor=west] {yes} (read);
  % \draw[arrow] (while) -- node[anchor=east] {no} (stop);
  % \draw[arrow] (read) -- (send);
  % \draw[arrow] (send) -- (waitack);
  % \draw[arrow] (ackrx) -- (ack);
  % \draw[arrow] (ack) -- node[anchor=east] {no} (send);
  % \draw[arrow] (ack) -| node[anchor=west] {yes} ([xshift=+1cm]send.south east) |- node[anchor=west] {$b$++} (while);
  % \draw[darrow] (waitack) -- node[anchor=west] {Ack received} (ackrx);
\end{tikzpicture}
\caption{Block diagram of a dual-connected device, an LTE eNB and a mmWave eNB}
\label{fig:mcdevice}
\end{figure}

An \texttt{McUeNetDevice} connects to an LTE eNB and to a mmWave eNB. In a dual connectivity scenario, the UE connects to the closest LTE cell, which in turn selects the mmWave cell with the highest SNR for initial connection of the UE mmWave stack. 
The mmWave eNB RRC does not interact with the core network when dealing with dual-connected devices, and only the LTE RRC handles the bearer creation and related protocol exchanges. The general scheme of an \texttt{McUeNetDevice}, an LTE BS and a mmWave eNB is shown in Fig.~\ref{fig:mcdevice}. In the dual connectivity setup the PDCP layer in the eNB relies on the X2 interface to transmit packets to the remote RLC layers in the mmWave eNB. In the UE, instead, the PDCP is directly connected to both LTE and mmWave RLCs.
The \texttt{McUeNetDevice} can be used also to simulate the handover between LTE and mmWave cells. Indeed, it can be configured to use a single RRC layer, which can be interfaced with PHY, MAC and RLC layers of both radio access technologies. 

Moreover, we extended the mmWave simulator classes in order to support handover and dual connectivity, by (i) providing SNR measurements in each eNB from all the UEs in the scenario; (ii) automatically configuring the UE protocol stack and PHY parameters after the connection to a new mmWave eNB; (iii) allowing disconnection of a UE from an eNB during handover; (iv) adding non contention based random access for initial access during handovers. The methods that allow the LTE macrocell to act as a coordinator in a dual connectivity scenario are implemented in the LTE eNB RRC layer. The LTE BS is thus capable of collecting measures from other mmWave eNBs, triggering handovers between mmWave cells or RAT switching. 

The mmWave simulator relies on a helper class \\(\texttt{MmWaveHelper}) which sets all the simulation parameters and objects needed. A \texttt{SpectrumChannel} models the mmWave channel. In order to support also communications on the LTE radio access network, the helper is extended with two additional \texttt{SpectrumChannel} objects, for LTE uplink and downlink channels, that can be characterized with different models of pathloss and are interfaced with the LTE physical layer. Moreover, we added methods to install in the simulation LTE eNBs, UEs and dual connected devices. The LTE eNBs can be characterized with their own scheduling and handover algorithms. 

X2 interfaces are created between all the eNBs, either LTE or mmWave. These links are modeled as ns--3 \texttt{PointToPoint} links, therefore it is possible to set the delay, data rate and Maximum Transfer Unit (MTU) of connections between eNBs, and between each eNB, the MME and the SGW. 

Actually, in the current implementation of the LTE module for ns--3 the link between eNBs and MME is ideal, whereas in this paper we extended the implementation to allow a more realistic modeling. 
In particular, the \texttt{EpcMme} class of the default LTE module is modified into an application (\texttt{EpcMmeApplication}), which is installed on an ns--3 node (\texttt{MmeNode}). Then this application is interfaced with the S1-MME endpoint at the MME node. The latter is connected with a \texttt{PointToPoint} link to the other endpoint, in the eNB, thus, as for the X2 links, it is possible to introduce a propagation delay, a data rate and an MTU also for S1-MME links.
The new \texttt{EpcS1Mme} and \texttt{EpcS1Enb} classes receive SDUs from the MME and the eNB, respectively, and create PDUs that can be sent over S1-MME by encoding the Information Elements and adding the S1-AP header\footnote{{S1-AP is the protocol which runs on top of the S1-MME link.}}, as specified in~\cite{3gppS1ap}.

\begin{figure*}[t]
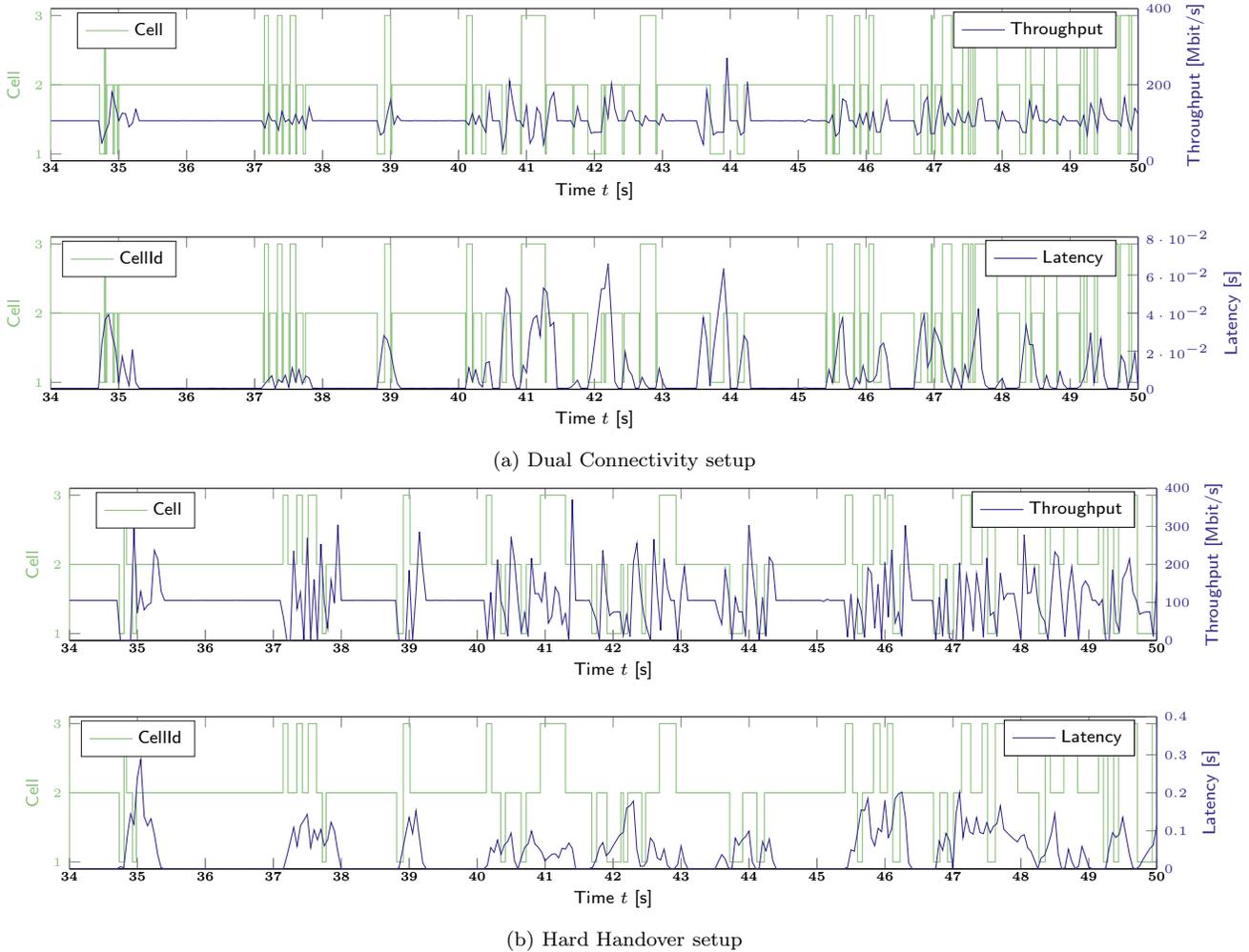

\sffamily\small
  \centering

\begin{subfigure}[t]{\textwidth}
	\centering
  \setlength\fwidth{0.9\textwidth}
  \setlength\fheight{0.3\textwidth}
  \input{./figures/dc_80_1ms_10MB_2ms.tex}
  \caption{Dual Connectivity setup}
  \label{fig:thexample}
\end{subfigure}
\begin{subfigure}[t]{\textwidth}
  \centering
  \setlength\fwidth{0.9\textwidth}
  \setlength\fheight{0.3\textwidth}
  \input{./figures/hh_80_1ms_10MB_2ms.tex}
  \caption{Hard Handover setup}
  \label{fig:latexample}
\end{subfigure}
\caption{Examples of simulation output: throughput and latency over time, over the cell to which the UE is connected. Cells 2 and 3 are the mmWave eNBs, cell 1 is the LTE eNB}
\label{fig:example}
\end{figure*}

\section{Metrics}\label{sec:metrics}
The main goal of this implementation is to provide a tool that can be used to compare the performance between a system using dual connectivity, with switching, and another where hard handover (HH) is used. 

The main benefit of an implementation of the LTE-5G tight integration in ns--3 is the possibility to study the performance of the system by considering a very high level of detail, with realistic interactions among the different parts of the network, from the radio channel modeling to the higher layer protocol message exchanges. Indeed, the comparison between the two systems can be affected by several parameters. The latency of X2 connections may have an impact on how quickly the LTE eNB detects that a UE is in outage with respect to the current mmWave link. The duration of the intervals during which a UE is performing the handover depends on the delay of the connection to the MME, and this affects the latency of data packet transmissions. By using real RRC messages one can account for a different delay and packet error rate in the exchange of control messages between the UE and the eNB. This also makes the parameter space very large, and a thorough simulation campaign to compare the two solutions is ongoing.

There are many metrics that can be computed in a simulation in ns--3 using the developed framework over LTE and mmWave modules. In particular, throughput and latency can be extracted at each layer, from PHY to transport. Packet losses during switching and handover events are also a metric of interest. Another issue that may arise on a high throughput mmWave link is bufferbloat at the RLC layer~\cite{mmNet}, and latency due to RLC retransmissions. Since ns--3 LTE and mmWave modules offer different RLC implementations, it will be possible to check whether these problems are present or not in the two systems.

The behavior of different transport protocols (UDP, TCP) can be tested, to check which of the two architectures offers higher resiliency in mobility scenarios. In particular it will be interesting to see how TCP reacts to the changes in the link in terms of latency and data rate. 

\begin{figure}[ht!]
	\centering
	\begin{tikzpicture}[font=\sffamily\large, scale=0.56, every node/.style={scale=0.56}]
		\centering

	    \node[anchor=south west,inner sep=0] (image) at (0,0) {\includegraphics[width=0.9\textwidth]{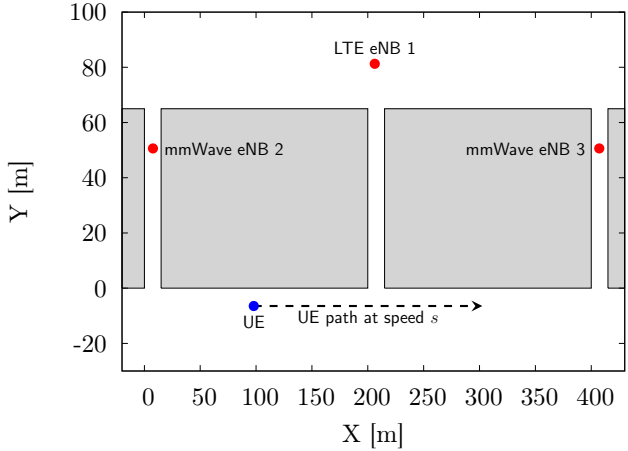}};
	    \begin{scope}[x={(image.south east)},y={(image.north west)}]
	        \filldraw[red,ultra thick] (0.23,0.645) circle (2pt);
	        \node[anchor=west] at (0.24,0.645) (mm1label) {mmWave eNB 2};
	        \filldraw[red,ultra thick] (0.894,0.645) circle (2pt);
	        \node[anchor=east] at (0.88,0.645) (mm2label) {mmWave eNB 3};
	        \filldraw[red,ultra thick] (0.56,0.825) circle (2pt);
	        \node[anchor=south] at (0.56,0.835) (ltelabel) {LTE eNB 1};

	        \draw[sarrow] (0.38, 0.31) -- node[anchor=north] {UE path at speed $s$} (0.72, 0.31);

	        \filldraw[blue,ultra thick] (0.38, 0.31) circle (2pt);
	        \node[anchor=north] at (0.38, 0.3) (ltelabel) {UE};
	    \end{scope}		
	\end{tikzpicture}
	\caption{Simulation scenario. The grey rectangles are buildings}
	\label{fig:scenario}
\end{figure}

\begin{table}[ht!]
  \centering
  \begin{tabular}{@{}ll@{}}
  	\toprule
    Parameter & Value \\
    \midrule
    Outage threshold & $- 5$ dB \\
    mmWave carrier frequency & 28 GHz \\
    mmWave bandwidth & 1 GHz \\
    LTE carrier frequency (DL) & 2.1 GHz \\
    LTE bandwidth & 20 MHz \\
    X2 link latency $D_{X2}$  & 1 ms \\
    RLC AM buffer size $B_{RLC}$ & 10 MB \\
    S1-MME link latency & 10 ms \\
    UDP packet size & 1024 byte \\
    UDP packet interarrival & 80 $\mu$s \\
    UE speed $s$ & 2 m/s along the x axis (Fig.~\ref{fig:scenario}) \\
    Iterations & $N=10$ \\ \bottomrule
  \end{tabular}
  \caption{Simulation parameters}
  \label{table:params}
\end{table}

As an example of metrics that can be extracted from the simulator, we present in this paper some early results from a first set of simulations. It is possible to directly compare a simulation with dual connectivity and a simulation with hard handover because by using the same parameters for the random number generation the channel varies in the same way. Table~\ref{table:params} contains some parameters of the simulations, and the scenario is shown in Fig.~\ref{fig:scenario}. The traffic is generated at a rate of 102.4 Mbit/s at the UDP transport layer, and sent from a remote server to the UE. The UE moves along the x axis at $s = 2$~m/s. An example of simulation output is provided in Fig.~\ref{fig:example}, which shows the throughput and latency at the PDCP layer over time. The UE moves from coordinates (100, -5) to (300, -5), and while moving may experience outage from both mmWave cells (thus the connection changes to LTE RAT either with handover or switching) or perform a handover to the mmWave cell with higher SNR. A hysteresis of 3 dB is accounted for when considering which is the best mmWave cell, and the current cell over time is shown by the green line. 

The average values of the throughput and latency, over a set of $N = 10$ simulations for this particular setup, are given in Table~\ref{table:av}. It can be seen that the latency of the dual connectivity solution is smaller than that of hard handover. This is due to the fact that the switch is much faster than the handover, therefore packets do not have to be buffered before being transmitted. This behavior is shown also in Fig.~\ref{fig:example}. For example, at $t=34$~s there is an outage event, i.e., the connection falls back to LTE cell 1, then the UE returns to a mmWave connection to mmWave cell 3, and finally it connects to mmWave cell 2. With the DC solution, the latency never exceeds 40 ms, while the HH latency exhibits a spike of 287 ms. 

The average PDCP throughput of DC is also higher than that of HH, showing that dual connectivity suffers fewer packet losses.

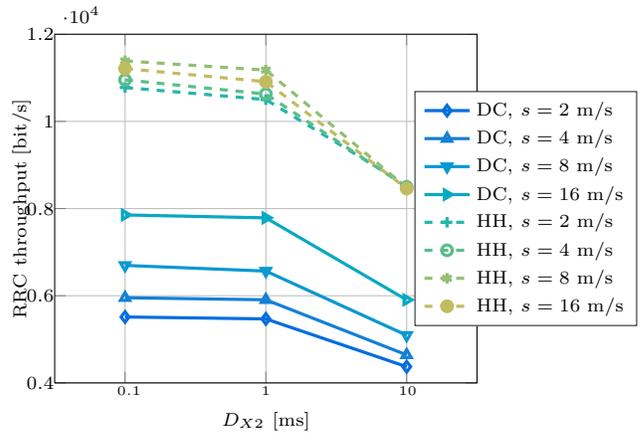
\begin{figure}[t!]
  \centering
  \setlength\fwidth{0.7\columnwidth}
  \setlength\fheight{0.55\columnwidth}
\definecolor{mycolor1}{rgb}{0.01651,0.42660,0.87863}%
\definecolor{mycolor2}{rgb}{0.14527,0.70976,0.66463}%
\definecolor{mycolor3}{rgb}{0.07794,0.50399,0.83837}%
\definecolor{mycolor4}{rgb}{0.34817,0.74243,0.54727}%
\definecolor{mycolor5}{rgb}{0.03434,0.59658,0.81985}%
\definecolor{mycolor6}{rgb}{0.57086,0.74852,0.44939}%
\definecolor{mycolor7}{rgb}{0.02666,0.66420,0.76072}%
\definecolor{mycolor8}{rgb}{0.75249,0.73840,0.37681}%
\begin{tikzpicture}
\tikzstyle{every node}=[font=\scriptsize]
\pgfplotsset{every x tick label/.append style={font=\tiny, yshift=0.5ex}}

\begin{axis}[%
width=0.951\fwidth,
height=\fheight,
at={(0\fwidth,0\fheight)},
scale only axis,
xmin=0.5,
xmax=3.5,
xlabel={$D_{X2}$ [ms]},
xtick={1,2,3},
xticklabels={$0.1$, $1$, $10$},
xticklabel style={text width=6, align=center}, xmajorgrids,
ymin=4000,
ymax=12000,
ylabel={RRC throughput [bit/s]},
ymajorgrids, ylabel shift = -5 pt, yticklabel shift = -3 pt,
axis background/.style={fill=white},
legend columns=1,
legend style={legend cell align=left,align=left,draw=white!15!black,at={(0.81\fwidth,0.5\fheight)},anchor=west}
]
\addplot [color=mycolor1,solid,line width=1.2pt,mark=diamond,mark options={solid}]
  table[row sep=crcr]{%
1	5512.66954068819\\
2	5468.48354158857\\
3	4371.5498896224\\
};
\addlegendentry{DC, $s =2$~m/s};

\addplot [color=mycolor3,solid,line width=1.2pt,mark=triangle,mark options={solid}]
  table[row sep=crcr]{%
1	5954.44870756334\\
2	5907.32793087574\\
3	4642.78093343619\\
};
\addlegendentry{DC, $s =4$~m/s};

\addplot [color=mycolor5,solid,line width=1.2pt,mark=triangle,mark options={solid,rotate=180}]
  table[row sep=crcr]{%
1	6696.08781924404\\
2	6565.80469566959\\
3	5092.69977101461\\
};
\addlegendentry{DC, $s =8$~m/s};

\addplot [color=mycolor7,solid,line width=1.2pt,mark=triangle,mark options={solid,rotate=270}]
  table[row sep=crcr]{%
1	7853.62177680839\\
2	7786.73661774196\\
3	5905.27671229914\\
};
\addlegendentry{DC, $s =16$~m/s};

\addplot [color=mycolor2,dashed,line width=1.2pt,mark=+,mark options={solid}]
  table[row sep=crcr]{%
1	10777.6639054363\\
2	10499.8802450844\\
3	8534.12585464779\\
};
\addlegendentry{HH, $s =2$~m/s};

\addplot [color=mycolor4,dashed,line width=1.2pt,mark=o,mark options={solid}]
  table[row sep=crcr]{%
1	10952.9507534999\\
2	10627.6201948363\\
3	8495.07202947351\\
};
\addlegendentry{HH, $s =4$~m/s};

\addplot [color=mycolor6,dashed,line width=1.2pt,mark=asterisk,mark options={solid}]
  table[row sep=crcr]{%
1	11388.2295066441\\
2	11176.7192657598\\
3	8481.97753811186\\
};
\addlegendentry{HH, $s =8$~m/s};

\addplot [color=mycolor8,dashed,line width=1.2pt,mark=*,mark options={solid}]
  table[row sep=crcr]{%
1	11211.5663381013\\
2	10910.0671826381\\
3	8459.37868907288\\
};
\addlegendentry{HH, $s =16$~m/s};

\end{axis}
\end{tikzpicture}%
  \caption{Average RRC traffic per user for DC and HH}
  \label{fig:rrc}
\end{figure}

Another example of metric is shown in Fig.~\ref{fig:rrc}, where we present the average traffic per user generated by the RRC layer for different values of the X2 latency $D_{X2} \in \{0.1, 1, 10\}$ ms and for different UE speed $s\in\{2,4,8,16\}$~m/s. It can be seen that the dual connectivity solution, with fast switching, allows to reduce the signaling traffic per user: this reduces the overhead, given a certain number of UEs, or, given the same amount of control overhead, it allows to scale to a larger number of UEs.

\begin{table}[t!]
  \centering
  \begin{tabular}{@{}lll@{}}
    \toprule
    & PDCP Throughput & RLC Latency \\ \midrule
    Dual Connectivity & $106.70$ Mbit/s & $5.1$ ms \\ 
    Hard Handover & $104.98$ Mbit/s & $18.1$ ms \\ 
    \bottomrule
  \end{tabular}
  \caption{Throughput and latency for the two setups, for the parameters in Table~\ref{table:params}, average over $N=10$ simulations}
  \label{table:av}
\end{table}

\section{Conclusions}

In this paper we discussed two possible ways to integrate 5G and LTE networks in order to improve the reliability of next generation mobile networks. We also presented the implementation of a simulation framework that can be used to assess the performance of such systems, integrated in ns--3, and showed that the level of detail of the simulation that can be carried out with such a tool makes it possible to understand and evaluate which is the best solution among dual connectivity with switching and hard handover. We showed some early results, for a particular choice of parameters, as an example of a possible simulation output. A more detailed
description of the new software modules and a more comprehensive set of preliminary results can be found in~\cite{masterthesis}. The application of the proposed framework to extensive simulation campaigns to fully characterize performance trends and to gain key insights for system design is left for future work.

\bibliographystyle{abbrv}
\bibliography{sigproc}  % sigproc.bib is the name of the Bibliography in this case
% ACM needs 'a single self-contained file'!

% That's all folks!
\end{document}